# Edge-based Collision Avoidance for Vehicles and Vulnerable Users

M. Malinverno[1], G. Avino[1], C. Casetti[1,4], C. F. Chiasserini[1,2,4], F. Malandrino[2,1,4], S. Scarpina[3]
1: Politecnico di Torino, Turin, Italy
2: CNR-IEIIT, Turin, Italy
3: TIM, Turin, Italy
4: CNIT, Italy

## Abstract

Collision avoidance is one of the most promising applications for vehicular networks, dramatically improving the safety of the vehicles that support it. In this paper, we investigate how it can be extended to benefit vulnerable users, e.g., pedestrians and bicycles, equipped with a smartphone. We argue that, owing to the reduced capabilities of smartphones compared to vehicular on-board units, traditional distributed approaches are not viable, and that multi-access edge computing (MEC) support is needed. Thus, we propose a MEC-based collision avoidance system, discussing its architecture and evaluating its performance. We find that, thanks to MEC, we are able to extend the protection of collision avoidance, traditionally thought for vehicles, to vulnerable users without impacting its effectiveness or latency.

## 1. Introduction

Saying that driving is dangerous would be an understatement. The United States National Highway Traffic Safety Administration (NHTSA) reported over 37,000 traffic fatalities for 2017, and nowadays the World Health Organization (WHO) estimates 3,400 *daily* traffic-related deaths, 50% of which could be avoided with appropriate action [1]. Motivated by these ghastly figures, *safety* has emerged as a prominent application of vehicular networks. Among safety applications, the most popular – and, arguably, the most effective – is *collision avoidance*. The idea of collision avoidance is fairly simple: vehicles are equipped with an on-board unit (OBU) that periodically [2] (and anonymously [3]) broadcasts a *Basic Safety Message[1]* (BSM) containing the vehicle's position, direction, acceleration, and speed. The OBU uses the BSMs sent by other vehicles to assess whether they are set on a collision course; if this is the case, the vehicle can alert its driver and/or take immediate action, e.g., perform an emergency brake.

Collision avoidance systems are especially important in presence of obstacles, e.g., buildings, that prevent drivers/vehicles from timely realizing the danger. Their importance and relevance have been acknowledged by transportation regulators: in December 2016, the U.S. Department of Transportation (DOT) published a Notice of Proposed Rulemaking (NPRM) for vehicular communications [4]. The document proposes to establish a new Federal Motor Vehicle Safety Standard (FMVSS), No. 150, to make vehicular networking technology compulsory: 50% of newly-made vehicles will have to be equipped with such a technology in 2021, 75% in 2022, and 100% in 2023. An important part of the picture, however, is missing. As reported by the WHO [1], half of the traffic fatalities concern *vulnerable users*, such as pedestrians and bicycles. Such users cannot, obviously, carry an OBU, which puts them out of the scope of traditional collision avoidance systems. On the positive side, vulnerable users do often carry smartphones, equipped with all the sensors – most notably, GNSS and accelerometer – needed for collision avoidance. Our intuition is therefore to leverage smartphones to *integrate* vulnerable users within collision avoidance systems, thereby extending to them the associated safety benefits.

---

[1] Equivalently, the Cooperative Awareness Messages (CAMs) standardized by ETSI could be considered.

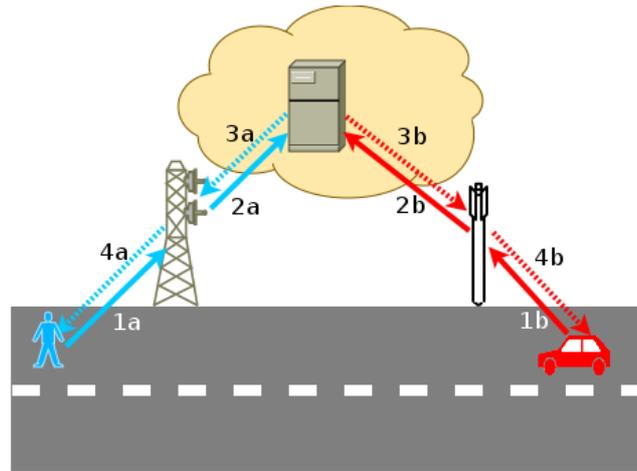

*Figure 1: Conceptual view of the proposed MEC-based architecture. POAs of different mobile networks collect BSMs from both vulnerable users (like the pedestrian in blue) and vehicles (like the car in red), as in steps 1a and 1b. POAs then convey the BSMs to a collision detection server (steps 2a and 2b), which processes them and establishes if there is a collision risk. If this is the case, alerts are conveyed to the appropriate POAs (steps 3a and 3b) and, hence, to the interested users (steps 4a and 4b).*

Smartphones, however, differ from OBUs in two key aspects. The first is their lack of support for network technologies like IEEE 802.11p/WAVE, very popular in vehicular networks. The second is represented by their computational power and energy limitations: constantly processing an endless flow of incoming BSMs would impose too much of a strain on the CPU and battery of a smartphone. Both these concerns can be addressed with the help of the multi-access edge computing (MEC) paradigm, where computation happens *within* the mobile network. In a MEC-based architecture such as the one exemplified in **Error! Reference source not found.**:

- vehicles and smartphones send their BSMs to the network infrastructure, i.e., to a point-of-access (POA) using a technology they support (e.g., CV2X for vehicles);
- BSMs are combined and processed within the network infrastructure;
- *alert* messages notifying the impeding collision are sent by the collision detector through the infrastructure to the entities set on a collision course, be them vehicles or vulnerable users.

Such an architecture adheres to the spirit of the MEC paradigm, as well as to the letter of its name: it is *multi-access*, as it integrates POAs using different technologies (including IEEE 802.11p RSUs and C-V2X base stations) and it leverages *edge computing*, thus relieving smartphones from the task of processing the BSMs.

The remainder of this paper is organized as follows. We begin by discussing related work in Sec. 2. Then Sec. 3 describes and discusses our system architecture, along with the underlying choices and trade-offs. Sec. 4 is devoted to the collision detection algorithm we use and its main parameters, while Secs. 5 and 6 present, respectively, our reference scenario and performance evaluation results. Finally, Sec. 7 concludes the paper and discusses the main research questions that still remain open.

## 2. Related work

Several works address safety applications in the automotive domain, including [5], [6] and [7]. In particular, [6] proposes a collision avoidance system for pedestrians, which exploits the pedestrians' smartphone to get information on, e.g., position and speed. In [7], Bazzi et al. analyze the performance of safety applications using the 3GPP LTE-V2V and the IEEE 802.11p technologies, with no mobile network infrastructure support.

Our solution includes a trajectory-based collision detection system, using a state-of-the art algorithm that we enhanced to match our needs. The basics of the algorithm have been used, with a different flavor, in [8], which presents a top-down and specification-driven design of an adaptive, peer-to-peer collision alert system. The performance of that version of the collision avoidance algorithm has been evaluated in [9]. With respect to [8] and [9], we have significantly enhanced both the algorithm and its parameters; furthermore, the system architecture has been extended in order to support distributed collision detection as well as the centralized one.

An overview of the main aspects and of vehicular network architectures and research issues can be found in [10]. A very good survey on the strengths and weaknesses of LTE as an enabler of vehicular communications is [11], where Araniti et al. extend some of the standardized safety messages we use in this work.

## 3. System architecture

The MEC-based system architecture we propose includes, as depicted in **Error! Reference source not found.**, three main entities: users (namely, vehicles and vulnerable users), POAs (of different technologies), and collision detection servers. These entities are described and discussed below.

### A. Users

In general, the *user* of a collision avoidance system is anything that could cause or be involved in a collision. This includes *vehicles* of different type and size, equipped with an OBU, and *vulnerable users*, such as pedestrian and bicycles, using smartphones *in lieu* of OBUs.

Regardless of their nature, users have to periodically broadcast BSMs, including the following information [12]:

- position, speed, and heading;
- lateral and vertical acceleration;
- vehicle length and width.

BSMs are sent in broadcast, so users may – but are not required to – listen to incoming BSMs and run their own collision detection algorithms. Vehicles are more likely to do so, while vulnerable users will probably rely only on the alerts issued by the collision detector.

What is done upon receiving an alert depends on the individual users. Vehicles can display a warning to their driver or perform an emergency braking. Smartphones can notify their owners through any combination of sound, vibration, and on-screen warning, disrupting any other activity in progress, such as music playback.

### B. POAs and network technologies

As mentioned, our architecture requires an *infrastructure*, i.e., a set of POAs connected with each other. This serves a twofold purpose: first, it makes it possible to integrate different technologies, thereby fulfilling the *multi-access* part of the MEC paradigm; second, it widens the set of BSMs available for collision detection, including those coming from users that are not in line-of-sight with each other.

The role of POAs is fairly straightforward, and mostly involves passing along BSMs from the users to the collision detector and collision alerts from the detector to the affected users. It is also interesting to remark that none of the POAs will *solely* be devoted to collision avoidance, or safety application, e.g., 802.11p RSUs support both safety and entertainment applications, e.g., on-board video.

### C. Collision detectors

Collision detectors are computing entities – physical servers, virtual machines, or containers – that run a collision detection algorithm. In general, collision detection takes as an input the BSMs generated by the vehicles or vulnerable users and forwarded by POAs, process them in order to establish whether any two are set on a collision course, and emit collision alerts if this is the case. Their basic internal architecture is described in Figure 2, while Sec. 4 presents the details of the collision detection algorithm we employ.

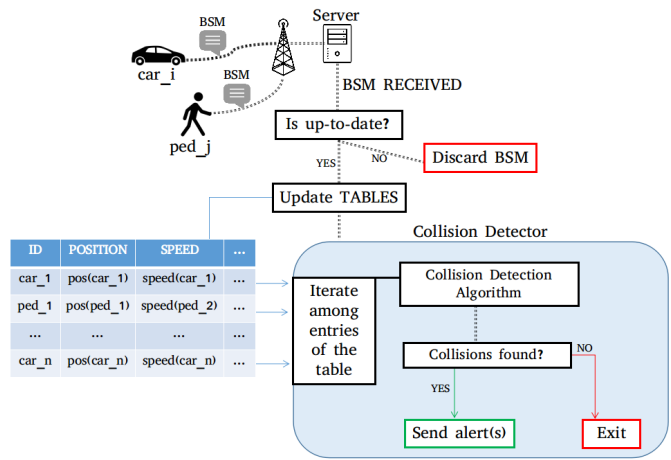

*Figure 2: Basic internal architecture of a collision detector. Incoming BSMs are first checked for obsolescence, and too-old BSMs are discarded. Then, the information included in the current BSM is checked against the elements of a table containing the position and speed of other vehicles that recently sent a BSM. If two vehicles are found to be set on a collision course, then the appropriate alerts are generated and sent. Finally, the table is updated with information from the newly-received BSM.*

The MEC paradigm offers significant flexibility in the *number* and the *placement* of collision detectors. Indeed, we can deploy multiple detectors at different positions in the network, from individual POAs to datacenters close to the network core. Choosing the location and number of collision detectors involves a trade-off between latency and effectiveness: detectors located closer to the vehicles, e.g., at individual POAs, result in shorter delays for both BSMs and alerts; furthermore, since they control a smaller area and, thus, have fewer BSMs to process, they can make faster decisions. On the other hand, detectors closer to the network core can receive data from more POAs covering a wider area, and, thus, detect more collisions (albeit with higher latency). Finally, if multiple detectors are deployed, they need to coordinate with each other in order to avoid that users receive repeated, or even contradictory, alerts.

## 4. Collision detection algorithm

Our collision detection algorithm, inspired by [8], is presented in Figure 3. It runs every time a new BSM is received and takes as an input the initial position $\vec{x}_0$, speed $\vec{v}$, and acceleration $\vec{a}$ contained in the newly-received BSM, as well as a set $\mathcal{B}$ including other entities that have recently sent a BSM. Note that, to reduce the processing burden, entities in $\mathcal{B}$ with which collision would be impossible, i.e., they are too far away or the mutual distance is growing in time, are filtered out. Also, for the sake of readability, the pseudocode in Figure 3 does not account for acceleration information, which is however used in our system to precisely predict the future positions.

The algorithm first initializes the set $\mathcal{C}$ of nodes with which the current entity could collide (line 1) and estimates the position of the sender of the newly-received BSM (tagged entity) at the future time instants. Then, the algorithm computes the position of each entity $b \in \mathcal{B}$ that recently sent a BSM (line 4) and vector $\vec{d}(t)$ representing the component-wise difference between the positions of the two entities (line 5). Note that the modulo of such a vector corresponds to the Euclidean distance. In line 6, we compute the square of such a distance $D(t)$ so as to simplify the subsequent computations. In the simplified version of the algorithm here shown, this calculation is relatively easy. In the complete version of the algorithm, which accounts for acceleration information, this part requires solving a 4$^{th}$ grade equation.

Since we are interested in the minimum value of $D(t)$, in line 7 we compute $t^*$, defined as the time instant at which the distance between the two entities is minimum. If $t^* < 0$, the two entities are getting farther apart, and no collision is going to happen; similarly, if $t^*$ is greater than a threshold $t2c_t$ (time to collision threshold), we are sure no collision will happen before $t2c_t$. In both cases, no action is

**Require:** $\vec{x}_0, \vec{v}, \mathcal{B}$
1: $\mathcal{C} \leftarrow \emptyset$
2: $\vec{x}(t) \leftarrow \vec{x}_0 + \vec{v}t$
3: **for all** $b \in \mathcal{B}$ **do**
4:  $\vec{x^b}(t) \leftarrow \vec{x_0^b} + \vec{v^b} \cdot t$
5:  $\vec{d}(t) \leftarrow \vec{x}(t) - \vec{x^b}(t)$
6:  $D(t) := |\vec{d}(t)|^2 \leftarrow (\vec{v} - \vec{v^b}) \cdot (\vec{v} - \vec{v^b})t^2 + 2(\vec{x_0} - \vec{x_0^b}) \cdot (\vec{v} - \vec{v^b})t + (\vec{x_0} - \vec{x_0^b}) \cdot (\vec{x_0} - \vec{x_0^b})$
7:  $t^* := t : \frac{d}{dt}D(t) = 0 \leftarrow \frac{-(\vec{x_0} - \vec{x_0^b}) \cdot (\vec{v} - \vec{v^b})}{|\vec{v} - \vec{v^b}|^2}$
8:  **if** $t^* < 0$ **or** $t^* > t2c_t$ **then**
9:    **continue**
10: $d^* \leftarrow \sqrt{D(t^*)}$
11: **if** $d^* \leq s2c_t$ **then**
12:   $\mathcal{C} \leftarrow \mathcal{C} \cup \{b\}$
13: **return** $\mathcal{C}$

*Figure 3: Collision detection algorithm.*

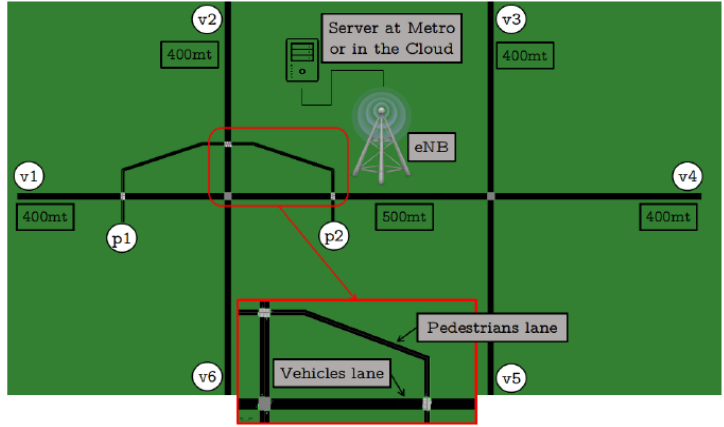

*Figure 4: Reference topology (SUMO screenshot).*

warranted (line 8). If, instead, $t^*$ is between 0 and $t2c_t$, then in line 11 we compute the minimum distance $d^*$ between the two entities, i.e., the distance at which they will be at time $t^*$. Such a distance is then compared against a minimum threshold $s2c_t$ (space to collision): if $d^*$ is lower, then a collision is deemed likely and $b$ is added to set $\mathcal{C}$ (line 12), else the algorithm moves to the next entity to process. Once all entities in $\mathcal{B}$ have been processed, the algorithm returns the set $\mathcal{C}$ of those entities with which the tagged one is on a collision course. If the set $\mathcal{C}$ is empty, then no action is taken; otherwise, an alert message is sent to the tagged entity as well as to all entities in set $\mathcal{C}$.

The $t2c_t$ and $s2c_t$ parameters depend on the individual scenario under consideration, and different values thereof result in different trade-offs between collision detection false positives and false negatives.

Accounting for acceleration as well as speed can significantly improve the accuracy of collision detection, especially when the speed of vehicles changes rapidly over time. On the negative side, accounting for acceleration values makes computations much more complex, especially solving the equation in line 7. Therefore, we run the version of the algorithm presented in **Error! Reference source not found.** if the vehicles report low acceleration, and the acceleration-aware version otherwise.

## 5. Reference scenario and simulations

In the following, we describe the reference scenario considered for our performance evaluation, the simulation tools we employ, and the metrics (key performance indicators, KPIs) we evaluate.

### A. Reference scenario and simulation tools

Our reference topology, depicted in Figure 4, is an urban area composed of three roads, crossing at two intersections, a pedestrian lane and three pedestrian crossings. Vehicles and pedestrians move throughout the topology, and all of them are covered by the cellular infrastructure (namely, an LTE eNB located at the center of the topology). Vehicles are equipped with onboard units (OBUs) for cellular vehicle-to-infrastructure (C-V2I) communications, whereas pedestrians carry a smartphone with cellular connectivity. Both periodically send BSMs toward the collision avoidance application server. The collision detection server is located, in pure MEC fashion, at a metro-level node within the cellular core network, incurring a latency of 5 ms. Note that, although our simulations consider a simple topology with straight roads, our algorithm is more general and – by considering vehicle acceleration – also works for curved roads.

New vehicles are generated at each of the ingress points v1-v6, following a Poisson distribution with rate of 0.7 vehicles per second, while pedestrians are generated at points p1-p2, still with a Poisson distribution, but rate equal to 0.3. Such rates have been chosen in order to guarantee that, on the one hand, there are enough entities in the topology to generate collisions and, on the other hand, that speeds do not become unrealistically slow due to congestion. The initial speed at which vehicles and pedestrians move coincides with their maximum speed, 13.89 m/s (50 km/h) and 2 m/s respectively. Vehicles never turn, e.g., vehicles generated at v2 always go towards v6. For vehicles, the BSM generation follows the dynamic scheme standardized by ETSI [12] – a mechanism designed for vehicular applications and specifically aiming at reducing the traffic load due to BSMs[2]. The maximum frequency is 10 Hz, while the minimum is 1 Hz. Consequently, the periodicity of BSM depends on the vehicle's speed, position, and heading angle: the higher the variation of one of these three parameters is, the higher the beaconing frequency will be. For pedestrians instead, given their low speed, dynamic BSM generation is not applied and their beaconing rate is capped to 1 Hz.

To avoid making decisions based upon outdated information, BSMs are discarded if they are older than 0.8 s, as highlighted in Figure 2. The $t2c_t$ and $s2c_t$ parameters for BSMs coming from vehicles are set to 10 s and 5 m, respectively, while for pedestrians these values are set to 5 s and 2 m, respectively. Such figures are based on a sensitivity analysis, which we omit for brevity. For what concerns the GPS positioning error, the target value for Collision Avoidance services is set to less than 1 m [13]. This value can be achieved through modern positioning systems, by exploiting the fusion of data coming from multiple sensors (e.g., GPS receiver as well as automotive radar [14, Fig. 7]).

We simulate this scenario using the SimuLTE-Veins simulator, which in turn is based on the OMNeT++ network simulator and the SUMO mobility simulator; both are best-in-class, open-source, tools. It is worth noticing that SUMO is, by default, a collision-free simulator, i.e., vehicles always brake in time to avoid colliding. In order to obtain collisions, we have to tweak its parameters, specifically, setting vehicle deceleration to zero and placing always-green traffic lights at the crossings.

Finally, we model the vehicle reaction to an alert as follows. Based on [9], we divide the interval between the time at which an alert is issued and the time at which the collision occurs in three sub-intervals, as summarized in Figure 5.

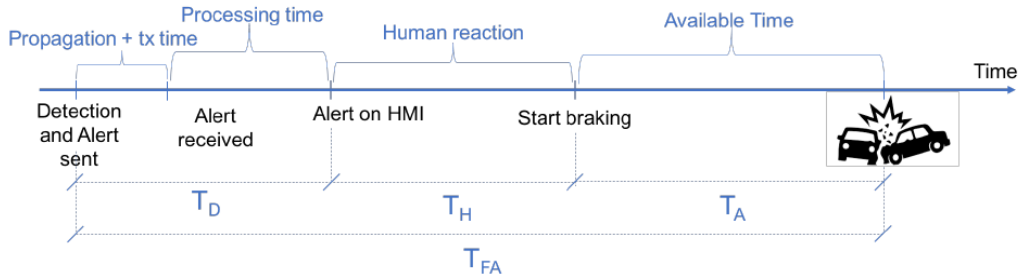

Figure 5: What occurs between the detection of a collision and the (potential) collision itself. The alert is sent on the wireless medium, received, and processed by the vehicle ($T_D$), then the human driver has to react and start braking ($T_H$); finally, the car decelerates and stops ($T_A$). If the collision detection service is intended for autonomous vehicles, then the alert triggers automated braking and $T_H=0$.

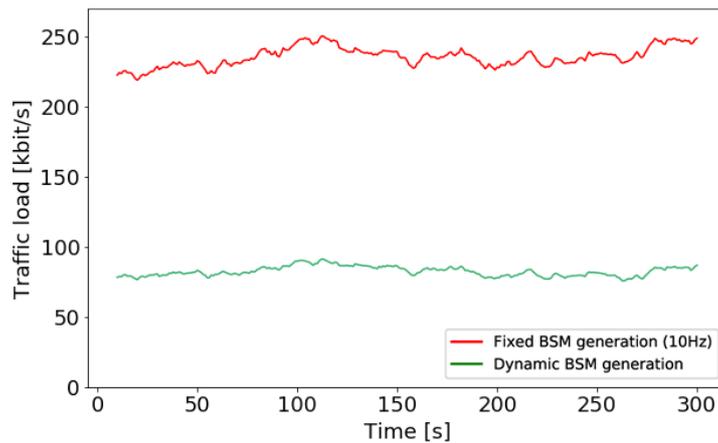

Figure 6: Gain in terms of traffic load due to BSM dynamic generation (green curve). This plot has been obtained through simulations with an average of 60 vehicles travelling over the considered geographical area. Under fixed BSM generation (red curve), the vehicles send messages with a frequency of 10 Hz.

---

[2] The dynamic BSM mechanism is not to be confused with the DCC scheme.

The first interval is indicated as $T_D$ and it includes the transmission and propagation time of the alert message (set to 5 ms in our scenario), as well as the processing time at the vehicle. In next-generation vehicles, the latter time is estimated to be equal to 400 ms [15]. After the processing time, an alert is displayed to the human driver, who then must react to it; such a reaction time is indicated as $T_H$ and can be conservatively estimated at one second, or at zero for autonomous vehicles and vehicles equipped with Advanced driver-assistance systems (ADAS). Finally, there is the actual braking time $T_A$, which depends on the vehicle type and initial speed; a collision is avoided if $T_A$ is sufficient to the vehicles involved to stop before they come in contact.

### B. KPIs of interest

In general, the evaluation of a collision avoidance system focuses on two main aspects: (i) how good it is at *detecting* collisions, i.e., assessing, based on the received BSMs, whether two users are set on a collision course, and (ii) how effective it is at *avoiding* collisions, i.e., if alerts are issued and delivered in time to prevent a collision.

For collision detection performance, we leverage the familiar false positive and false negative metrics, where:

- false positives correspond to pairs of users (vehicles or pedestrians) for which a collision alert is issued, but do not actually collide;
- false negatives correspond to pairs of users for which no alert is issued but do collide.

It is worth stressing that false positives can be as harmful as false negatives to the effectiveness of the collision avoidance system; indeed, too many false positives annoy the drivers and increase the likelihood they will not react appropriately to future warnings or disable the system altogether.

As described above, the effectiveness of a collision avoidance system in *avoiding* those collisions that are properly detected has to do with the *timeliness* of alerts.

## 6. Performance evaluation

As a preliminary aspect, we are interested in assessing the impact of dynamic BSM generation on the traffic load. To this end, we compare dynamic BSM generation against a fixed, 10-Hz generation rate in Figure 6, which shows a reduction of the traffic load due to BSMs by over 30%. Importantly, such a reduction does not have any impact on the application performance: the MEC-based Collision Avoidance can successfully detect all vehicle collisions in both cases (the plot is omitted for brevity).

We now move to the effectiveness of our system at detecting and avoiding collisions, expressed through the KPIs described in Sec. B. We are interested in two distinct but complementary aspects, i.e., (i) the performance of collision detection and avoidance for vulnerable users, and (ii) the impact of using a MEC architecture on the performance for vehicular users.

### A. Performance for vulnerable users

Our system is able to detect 100% of the collisions between vehicles and vulnerable users. This excellent result is due to the detection algorithm presented in Sec. 4, and it further highlights how significant the benefits of extending collision avoidance protection to vulnerable users are.

We are also interested in studying the *safety margin* associated with collision detection. Intuitively, we would like collision detection systems to warn the vehicles not only in time to avoid the collision, as depicted in Figure 5, but with some room to spare. We quantify such a margin by considering the distance at which two entities (in this case, a vehicle and a pedestrian) find themselves after braking/stopping as a consequence of a collision alert. Figure 7 shows the distribution of such a distance; we can observe that it always exceeds five meters – a very good margin, even allowing for such mishaps as delayed notifications on mobile phones. Indeed, one may say that collisions involving pedestrians are easier to avoid once they are detected, owing to the pedestrians' lower speed, hence time required to stop.

### B. Impact of the MEC architecture on vehicular users

We now assess whether moving from a decentralized architecture based on IEEE 802.11p V2V communications, where each vehicle runs its own collision detector, to a MEC-based one affects the collision avoidance performance for vehicular users. It is worth to underline that the distributed approach introduces smaller delays with respect to the centralized one, since BSMs and alerts now imply just one-hop transmissions. Nevertheless, the difference in terms of message delivery delay between distributed and centralized implementation is not very significant, due to the fact that the MEC architecture adopted for the centralized implementation has itself very low delay. Given the fact that cellular connectivity between mobile users and infrastructure is well established, we assume a penetration rate equal to 1 for the cellular case. As far as IEEE 802.11p V2V is concerned, we consider instead a 0.5 penetration rate, which can be realistic in the future. The propagation model we use accounts for the nodes' transmission power, the antenna height (which clearly favors V2I communications), path loss, and Nakagami-m distributed fading. Furthermore, for each of the two technologies, we consider both line-of-sight (LOS) and non-line-of-sight (NLOS) propagation conditions, with the latter accounting for the extra attenuation due to the presence of buildings.

**Error! Reference source not found.**(left) compares the performance of the centralized and distributed approach in terms of fraction of vehicle-with-vehicle collisions that are:

- detected on time, hence avoided, or
- detected too late (an alert is sent, but too late to avoid the collision), or
- not detected at all.

Observe that, when all entities are in LOS, the MEC architecture outperforms the decentralized architecture, with the latter failing to detect 74.9% of the collisions. Indeed, the lower technology penetration rate penalizes the IEEE 802.11p V2V technology, since both the vehicles involved in a collision must be equipped with the proper communication interface to avoid a collision.

In the presence of buildings, the detection rate drops to 10% in the distributed case, while it is still over 90% in the centralized scenario. This suggests that in general the centralized approach is much more resilient to harsh propagation conditions and urban canyoning with respect to the distributed counterpart.

On the negative side, **Error! Reference source not found.**(right) shows that the MEC architecture is associated with a marginally higher rate of false positives.

Both plots are consistent with the fact that, owing to the more centralized network architecture and to the better coverage of the eNB compared to individual vehicles, a larger number of BSMs are conveyed to the collision detector. This allows it to detect more collisions, but also makes false positives more likely.

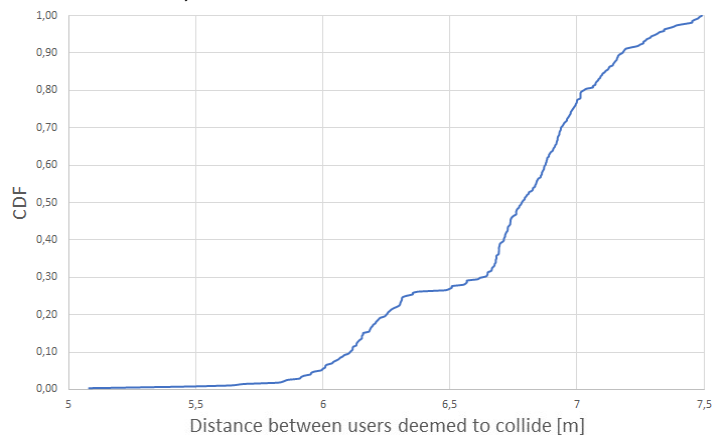

*Figure 7: Car-to-pedestrian collisions: distance between users deemed to collide.*

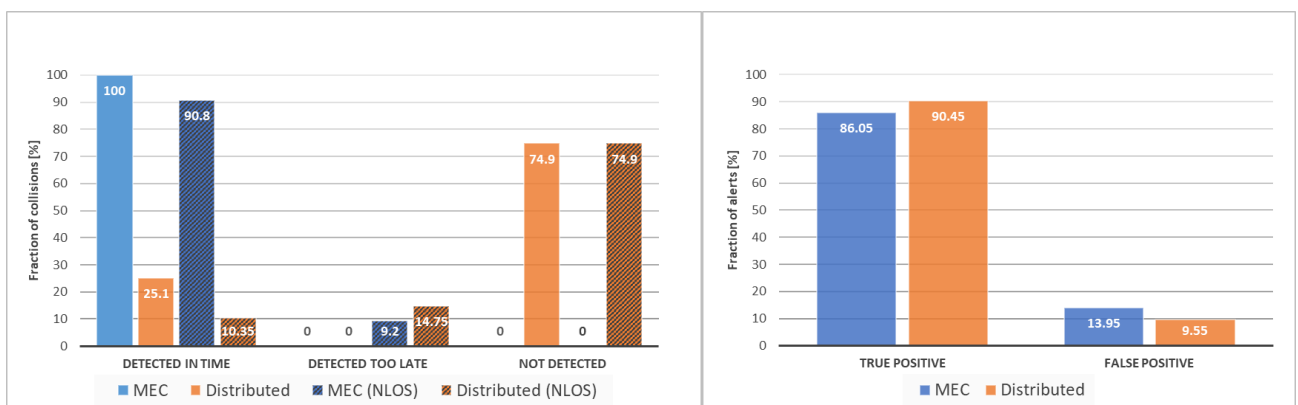

*Figure 8 Car-to-car collisions: fraction of detected collisions (left) and of true/false positives (right), for the centralized (MEC-based) and the decentralized architecture.*

# 7. Conclusion and future work

We endeavored to extend the protection of collision avoidance systems to vulnerable road users such as pedestrians and bikes. To this end, we proposed an architecture based on multi-access edge computing (MEC), where vehicles and vulnerable users can exploit different network technologies to send cooperative awareness messages (BSMs) to a centralized collision detector and receive collision alerts from it. Our performance evaluation showed that 100% of the collisions involving vulnerable users can be detected in time to be avoided, and that the performance of collision avoidance for vehicles is essentially the same as under a more traditional, decentralized architecture.

Future work will involve three main areas, namely, (i) improving the collision detection algorithm we use, adapting it to the MEC architecture, (ii) studying larger-scale scenarios featuring multiple collision detectors, and (iii) introducing a model for GPS positioning error.

# Acknowledgement

This work was partially supported by TIM through the research contract "Multi-access Edge Computing", and by the European Commission through the Horizon 2020 project 5G-CARMEN (grant agreement no. 825012).

# Authors

**Marco Malinverno** is a Ph.D. student with Politecnico di Torino, Turin, Italy. His research interests include vehicular networking and container-based solutions for service delivery. He is affiliated with the inter-disciplinary center CARS (Center for Automotive Research and Sustainable mobility).

**Giuseppe Avino** is a Ph.D. student with Politecnico di Torino, Turin, Italy. His research interests include vehicular networking, next-generation cellular networking, and container-based virtualization. He is involved in the EU H2020 project 5G-TRANSFORMER.

**Claudio Casetti** is an associate professor with Politecnico di Torino, Turin, Italy. His research interests include vehicular networks, 5G networks, transport and network protocols in wired networks, IEEE 802.11 WLAN. He is the chair of the Turin Urban Digital Mobility working group within the Smart Roads project fostered by the City of Turin.

**Carla Fabiana Chiasserini** is a full professor with Politecnico di Torino, Turin, Italy, and a research associate with CNR-IEIIT. Her research interests are mainly in the field of wireless and mobile networks, including 5G networks and vehicular networks. She is the Rector's Delegate for Alumni and Career Orientation, and is in the list of top Italian scientists kept by VIAAcademy. She is an IEEE fellow (class of 2018).

**Francesco Malandrino** is a researcher at CNR-IEIIT, Turin, Italy. His research interests include the architecture and management of wireless, cellular, and vehicular networks. His awards include a Fibonacci Fellowship from the Italian Ministry of Foreign Affairs. He is an IEEE senior member.

**Salvatore Scarpina** is a project manager at TIM, formerly known as Telecom Italia. He leads research projects in several areas, including smart cities, smart working, and 5G mobile networks. He is the TIM delegate at ETSI MEC (Multi-access Edge Computing) industry specification group and at Open Radio Access Network (O-RAN) Alliance.